# A High Performance Disturbance Observer

Emre Sariyildiz
School of Mechanical, Materials, Mechatronic, and Biomedical Engineering,
Faculty of Engineering and Information Sciences
University of Wollongong
Wollongong, NSW, 2522, Australia
emre@uow.edu.au

*Abstract*— This paper proposes a novel Disturbance Observer (DOb), termed the High-Performance Disturbance Observer (HPDOb), which achieves more accurate disturbance estimation compared to the conventional DOb, thereby delivering significant improvements in robustness and performance for motion control systems. The conventional and proposed DObs are analysed and synthesised in the discrete-time domain to fully capture their dynamic behaviours in real-world applications. This allows us to adjust the stability and performance of DOb-based digital robust motion control systems more effectively in practice. The conventional DOb is synthesised by assuming that the internal and external disturbances of a servo system does not change within the sampling period. Despite its simplicity, the constant disturbance model is impractical in many motion control applications. To address this issue, a more realistic disturbance model is utilised in the synthesis of the HPDOb, which is carried out in two steps. In the first step, a predictor is employed to estimate the disturbance change within the sampling period. The predictor can be implemented using the conventional DOb. In the second step, the predicted disturbance change is substituted into the high performance observer's dynamic model so that disturbances are estimated more accurately. Simulation results demonstrate that the proposed HPDOb significantly enhances the accuracy of disturbance estimation. This approach benefits a wide range of motion control applications by providing more accurate disturbance estimates to motion controllers, thereby enhancing both the robustness and performance of position control and sensorless force control systems.

*Keywords—Disturbance Observer, Reaction Force Observer, High-Performance Disturbance Estimation, and Robust Motion Control.*

## I. Introduction

Since its introduction by Ohnishi in 1983 [1], DOb has been widely adopted by motion control researchers, mainly for robust control applications [2–5]. Over the past three decades, DOb has been applied to a wide range of engineering problems, from robust position control of servo systems to interaction force estimation in robotics and time-delay compensation in communication systems [6–11]. In the robust control approach utilising a DOb, motion control systems achieve robustness by cancelling disturbances with their estimations in an inner-loop [12, 13]. To implement this intuitive robust control technique, different DOb synthesis methods have been proposed in the last three decades [14–19].

To synthesise a DOb, a dynamic model for the disturbance exerting on the servo system should be introduced [20]. Conventionally, a DOb is synthesised in the continuous-time domain under the assumption that the servo system is affected solely by constant disturbances [13]. There are two primary drawbacks in this conventional DOb synthesis approach. First, DOb-based robust motion control systems are always implemented by digital controllers such as computers and micro-controllers, and the dynamic behaviour of a digital DOb cannot be fully captured in the continuous-time domain [21–24]. As reported in [21–22], this leads to unexpected stability and performance problems in motion control applications. Second, the constant disturbance model employed in the conventional DOb synthesis is a very strict assumption in motion control, as servo systems are generally affected not only by constant disturbances but also by varying disturbances [25, 26]. This strict assumption restricts the accuracy of disturbance estimation, thereby limited robustness against disturbances and motion control performance in practice [27].

In pursuit of high-performance motion control, several analysis and synthesis methods have been proposed for digital robust motion controllers implemented using a DOb [28–32]. For instance, the control parameters of digital DObs are tuned employing optimal plant shaping algorithms, as discussed in [29–31], and the unexpected dynamic behaviours of digital DOb are explained in [32]. Besides, disturbance estimation accuracy is improved using different disturbance models in the DOb synthesis. While periodic disturbances are accurately estimated using the periodic DOb [33], a generalised DOb is developed to improve disturbance estimation accuracy through higher-order disturbance estimates [34]. Numerous studies have demonstrated that higher-order DObs can enhance disturbance estimation performance; however, they are also more susceptible to noise sensitivity [35–37]. Therefore, the trade-off between the robustness against disturbances and the controller's noise sensitivity becomes stricter when high-order DObs are employed in the robust motion controller synthesis [15, 19, 35]. Despite existing studies, further efforts are needed to improve digital DObs for enhanced robustness and estimation accuracy.

This paper presents a novel digital DOb that can provide high-performance in disturbance estimation. Compared to the conventional DOb, the proposed digital HPDOb enables the use of higher-order Taylor series expansions of the disturbance in the observer synthesis, thereby improving the accuracy of disturbance estimation. This can provide great benefits in motion control applications, e.g., boosting robustness against disturbances in position control and reducing sensorless force estimation error in interaction control. The validity of the proposed HPDOb is verified through simulations.

The remainder of the paper is structured as follows. Section II introduces the conventional DOb synthesis in the discrete–time domain. A simple yet effective design method is presented for robust motion control implementations. Section III proposes a novel DOb that can estimate disturbances more accurately than the conventional DOb. The validity of the proposed HPDOb is verified by simulations in Section IV. The paper is concluded in Section V.

## II. CONVENTIONAL DISTURBANCE OBSERVER

*a) Nominal System Model in the Discrete-time Domain:*

To synthesise a DOb-based robust motion controller, we need to introduce a nominal dynamic model for a servo system. Let us employ the following uncertain and nominal state space models in the DOb synthesis.

$$\dot{\mathbf{x}}(t) = \mathbf{A}_\mathbf{C}\mathbf{x}(t) + \mathbf{B}_\mathbf{C}u(t) - \mathbf{D}_\mathbf{C}\tau_d(t) \\ = \mathbf{A}_{\mathbf{Cn}}\mathbf{x}(t) + \mathbf{B}_{\mathbf{Cn}}u(t) - \mathbf{D}_{\mathbf{Cn}}\tau_{dn}(t) \quad (1)$$

where $\mathbf{A}_\mathbf{C} = \begin{bmatrix} 0 & 1 \\ 0 & b_m/J_m \end{bmatrix}$ and $\mathbf{A}_{\mathbf{Cn}} = \begin{bmatrix} 0 & 1 \\ 0 & b_{mn}/J_{mn} \end{bmatrix}$ represent the uncertain and nominal state matrices in which $J_m, J_{mn}, b_m$ and $b_{mn}$ are the uncertain and nominal inertiæ and viscous friction coefficients, respectively; $\mathbf{B}_\mathbf{C} = \begin{bmatrix} 0 \\ 1/J_m \end{bmatrix}$ and $\mathbf{B}_{\mathbf{Cn}} = \begin{bmatrix} 0 \\ 1/J_{mn} \end{bmatrix}$ are the uncertain and nominal control input vectors, respectively; $\mathbf{D}_\mathbf{C} = \begin{bmatrix} 0 \\ 1/J_m \end{bmatrix}$ and $\mathbf{D}_{\mathbf{Cn}} = \begin{bmatrix} 0 \\ 1/J_{mn} \end{bmatrix}$ are the uncertain and nominal disturbance input vectors, respectively; $\mathbf{x}(t) = \begin{bmatrix} q(t) & \dot{q}(t) \end{bmatrix}^T$ represents the state vector of the servo system in which $q(t)$ and $\dot{q}(t)$ are the position and velocity states, respectively; $u(t)$ represents the control input of the servo system; $\tau_d(t)$ represents external disturbances such as load and unmodeled dynamics; and $\tau_{dn}(t)$ represents a nominal disturbance variable described in Eq. (2).

$$\tau_{dn}(t) = \frac{1}{\mathbf{D}_{\mathbf{Cn}}^T \mathbf{D}_{\mathbf{Cn}}} \left( (\mathbf{A}_{\mathbf{Cn}} - \mathbf{A}_\mathbf{C})\mathbf{x}(t) + (\mathbf{B}_{\mathbf{Cn}} - \mathbf{B}_\mathbf{C})u(t) + \mathbf{D}_\mathbf{C}\tau_d(t) \right) \quad (2)$$

For a more detailed dynamic model of a servo system, the reader is referred to [12, 16].

Without any simplification, the state space model of a servo system can be similarly described in the discrete-time domain as follows:

$$\dot{\mathbf{x}}((k+1)T_s) = \mathbf{A}_\mathbf{D}\mathbf{x}(kT_s) + \mathbf{B}_\mathbf{D}u(kT_s) - \mathbf{\Pi}_\mathbf{D}(kT_s) \\ = \mathbf{A}_{\mathbf{Dn}}\mathbf{x}(kT_s) + \mathbf{B}_{\mathbf{Dn}}u(kT_s) - \mathbf{\Pi}_{\mathbf{Dn}}(kT_s) \quad (3)$$

where $\mathbf{A}_\mathbf{D} = e^{\mathbf{A}_\mathbf{C}T_s}$ and $\mathbf{A}_{\mathbf{Dn}} = e^{\mathbf{A}_{\mathbf{Cn}}T_s}$ are the uncertain and nominal discrete state matrices of the servo system, respectively; $\mathbf{B}_\mathbf{D} = \int_0^{T_s} e^{\mathbf{A}_\mathbf{C}\tau}\mathbf{B}_\mathbf{C}d\tau$ and $\mathbf{B}_{\mathbf{Dn}} = \int_0^{T_s} e^{\mathbf{A}_{\mathbf{Cn}}\tau}\mathbf{B}_{\mathbf{Cn}}d\tau$ are the uncertain and nominal discrete control input vectors, respectively; $\mathbf{\Pi}_\mathbf{D}(kT_s) = \int_0^{T_s} e^{\mathbf{A}_\mathbf{C}\tau}\mathbf{D}_\mathbf{C}\tau_d((k+1)T_s - \tau)d\tau$ and $\mathbf{\Pi}_{\mathbf{Dn}}(kT_s) = \int_0^{T_s} e^{\mathbf{A}_{\mathbf{Cn}}\tau}\mathbf{D}_{\mathbf{Cn}}\tau_{dn}((k+1)T_s - \tau)d\tau$ are the discrete disturbance vectors in the uncertain and nominal models, respectively; $\mathbf{x}(kT_s) = \begin{bmatrix} q(kT_s) & \dot{q}(kT_s) \end{bmatrix}^T$ is the state vector describing the position and velocity states of the servo system at $kT_s$ seconds; and $u(kT_s)$ is the control input at $kT_s$ seconds [38, 39].

*b) Conventional DOb Synthesis:*

Conventionally, constant disturbance model, e.g., $\dot{\tau}_{dn}(t) = 0$, is employed in the design of DOb-based robust motion control systems. To synthesise the conventional DOb in the discrete-time domain, let us assume that the nominal disturbance variable is constant within the sampling period, i.e., $\tau_{dn}((k+1)T_s) \cong \tau_{dn}(kT_s)$. Applying this assumption into the introduced nominal dynamic model yields

$$\mathbf{x}((k+1)T_s) = \mathbf{A}_{\mathbf{Dn}}\mathbf{x}(kT_s) + \mathbf{B}_{\mathbf{Dn}}u(kT_s) - \mathbf{D}_{\mathbf{Dn}}\tau_{dn}(kT_s) \quad (4)$$

where $\tau_{dn}(kT_s)$ represents the nominal disturbance variable at $kT_s$ seconds, and $\mathbf{D}_{\mathbf{Dn}} = \int_0^{T_s} e^{\mathbf{A}_{\mathbf{Cn}}\tau}\mathbf{D}_{\mathbf{Cn}}d\tau$ is the nominal disturbance input vector.

To estimate the nominal disturbance variable without using the derivative of the state vector, let us synthesise an observer for the auxiliary variable given by

$$z(kT_s) = \tau_{dn}(kT_s) + \mathbf{L}^T\mathbf{x}(kT_s) \quad (5)$$

where $z(kT_s) \in R$ is an auxiliary variable, and $\mathbf{L} \in R^2$ is the observer gain vector which will be tuned after deriving the dynamic model of the auxiliary variable observer [13].

By substituting the nominal state space model into Eq. (5), the dynamics of the auxiliary variable is derived as follows:

$$z((k+1)T_s) = (1 - \mathbf{L}^T\mathbf{D}_{\mathbf{Dn}})z(kT_s) + \mathbf{L}^T(\mathbf{A}_{\mathbf{Dn}} + \mathbf{D}_{\mathbf{Dn}}\mathbf{L}^T - \mathbf{I}_2)\mathbf{x}(kT_s) + \\ \mathbf{L}^T\mathbf{B}_{\mathbf{Dn}}u(kT_s) + \tau_{dn}((k+1)T_s) - \tau_{dn}(kT_s) \quad (6)$$

where $\mathbf{I}_2$ is a 2x2 identity matrix.

Since the nominal dynamic model of the servo system is utilised in deriving auxiliary variable dynamics, all parameters in Eq. (6) are known, except for the variation in the nominal disturbance variable. By neglecting the nominal disturbance variable's variation within the sampling period, the auxiliary variable observer can be synthesized as follows:

$$\hat{z}((k+1)T_s) = (1 - \mathbf{L}^T\mathbf{D}_{\mathbf{Dn}})\hat{z}(kT_s) + \mathbf{L}^T(\mathbf{A}_{\mathbf{Dn}} + \mathbf{D}_{\mathbf{Dn}}\mathbf{L}^T - \mathbf{I}_2)\mathbf{x}(kT_s) + \\ \mathbf{L}^T\mathbf{B}_{\mathbf{Dn}}u(kT_s) \quad (7)$$

where $\hat{z}(kT_s)$ represents the estimated $z(kT_s)$ at $kT_s$ seconds.

When the observer gain vector $\mathbf{L}$ is properly tuned, the estimated auxiliary variable converges to the exact one. Hence, the conventional DOb that estimates the nominal disturbance variable can be synthesised using Eqs. (7) and (8) as follows:

$$\hat{\tau}_{dn}(kT_s) = \hat{z}(kT_s) - \mathbf{L}^T\mathbf{x}(kT_s) \quad (8)$$

where $\hat{\tau}_{dn}(kT_s)$ represents the estimated $\tau_{dn}(kT_s)$ at $kT_s$ seconds.

*c) Conventional DOb Analysis:*

To tune the observer gain vector $\mathbf{L}$ for achieving good performance and robust stability, let us derive the dynamics of the auxiliary variable estimation error. Subtracting Eq. (7) from Eq. (6) yields

$$e_z((k+1)T_s) = (1-\mathbf{L}^T\mathbf{D_{Dn}})e_z(kT_s) + \Delta\tau_{dn}(kT_s) \quad (9)$$

where $e_z(kT_s) = z(kT_s) - \hat{z}(kT_s)$ is the auxiliary variable estimation error at $kT_s$ seconds, and $\Delta\tau_{dn}(kT_s) = \tau_{dn}((k+1)T_s) - \tau_{dn}(kT_s)$ is the variation of the nominal disturbance variable between $kT_s$ and $(k+1)T_s$ seconds.

As shown in Eq. (9), the conventional DOb is stable when the observer gain vector is tuned so that $|1-\mathbf{L}^T\mathbf{D_{Dn}}|<1$. When the nominal disturbance variable is constant, $\Delta\tau_{dn}(kT_s)$ is null and the auxiliary variable estimation error converges to zero as time progress. However, asymptotic stability cannot be achieved when the nominal disturbance variable is not fixed within a sampling period, and the upper bound of the estimation error is proportional to the variation of the nominal disturbance variable $\Delta\tau_{dn}(kT_s)$ [38, 39].

## III. A High-Performance Disturbance Observer

To enhance disturbance estimation accuracy, let us synthesise the HPDOb using the following two steps algorithm.

*Disturbance Predictor:*

Let us employ the auxiliary variable given in Eq. (10) in the predictor synthesis.

$$z_p(kT_s) = \tau_{dnp}(kT_s) + \mathbf{L_p}^T\mathbf{x}(kT_s) \quad (10)$$

where $z_p(kT_s) \in R$ is an auxiliary variable used in the predictor synthesis, $\tau_{dnp}(kT_s) \in R$ is the predicted nominal disturbance variable at $kT_s$ seconds, and $\mathbf{L_p} \in R^2$ is the gain vector of the predictor.

Substituting Eq. (4) into Eq. (10) yields

$$z_p((k+1)T_s) = (1-\mathbf{L_p}^T\mathbf{D_{Dn}})z_p(kT_s) + \mathbf{L_p}^T(\mathbf{A_{Dn}}+\mathbf{D_{Dn}}\mathbf{L_p}^T-\mathbf{I_2})\mathbf{x}(kT_s) + \\ \mathbf{L_p}^T\mathbf{B_{Dn}}u(kT_s) + \tau_{dnp}((k+1)T_s) - \tau_{dnp}(kT_s) \quad (11)$$

To estimate the auxiliary variable $z_p(kT_s)$, let us employ the following dynamic model which is derived by neglecting the variation of the nominal disturbance variable in Eq. (11).

$$\hat{z}_p((k+1)T_s) = (1-\mathbf{L_p}^T\mathbf{D_{Dn}})\hat{z}_p(kT_s) + \mathbf{L_p}^T(\mathbf{A_{Dn}}+\mathbf{D_{Dn}}\mathbf{L_p}^T-\mathbf{I_2})\mathbf{x}(kT_s) + \\ \mathbf{L_p}^T\mathbf{B_{Dn}}u(kT_s) \quad (12)$$

where $\hat{z}_p(kT_s)$ represents the estimated $z_p(kT_s)$ at $kT_s$ seconds.

Equations (6), (7), (11) and (12) show that the proposed predictor has the same dynamic model with the conventional DOb. Therefore, the stability of the predictor can be similarly obtained when the predictor gain vector $\mathbf{L_p}$ satisfies the condition $|1-\mathbf{L_p}^T\mathbf{D_{Dn}}|<1$. Hence, the nominal disturbance variable can be predicted using $\hat{\tau}_{dnp}(kT_s) = \hat{z}_p(kT_s) - \mathbf{L_p}^T\mathbf{x}(kT_s)$.

The predicted nominal disturbance variable can be used to estimate the derivatives of the disturbance as follows:

$$\dot{\hat{\tau}}_{dn}(kT_s) = (\hat{\tau}_{dnp}(kT_s) - \hat{\tau}_{dnp}((k-1)T_s))/T_s$$
$$\ddot{\hat{\tau}}_{dn}(kT_s) = (\hat{\tau}_{dnp}(kT_s) - 2\hat{\tau}_{dnp}((k-1)T_s) + \hat{\tau}_{dnp}((k-2)T_s))/T_s^2 \quad (13)$$
$$\vdots$$

The conventional DOb is synthesised using the zero-order Taylor series expansion of the nominal disturbance variable, i.e., $\tau_{dn}((k+1)T_s) = \tau_{dn}(kT_s)$. Equation (13) allows us to introduce more accurate disturbance models in the high-performance DOb synthesis. For example, the first and second order Taylor series expansions of the disturbance variable can be obtained using Eq. (13) as follows:

$$\hat{\tau}_{dn}((k+1)T_s) = \hat{\tau}_{dn}(kT_s) + \dot{\hat{\tau}}_{dn}(kT_s)T_s$$
$$\hat{\tau}_{dn}((k+1)T_s) = \hat{\tau}_{dn}(kT_s) + \dot{\hat{\tau}}_{dn}(kT_s)T_s + \frac{1}{2}\ddot{\hat{\tau}}_{dn}(kT_s)T_s^2 \quad (14)$$

*High-Performance Disturbance Observer:*

Let us now synthesise the HPDOb using the predicted disturbance variable and its derivatives at $kT_s$ seconds. The following auxiliary variable is used in the synthesis of the HPDOb.

$$z_o(kT_s) = \tau_{dn}(kT_s) + \mathbf{L_o}^T\mathbf{x}(kT_s) \quad (15)$$

where $z_o(kT_s) \in R$ is an auxiliary variable used in the synthesis of the HPDOb, and $\mathbf{L_o} \in R^2$ is a gain vector which is yet to be tuned.

The dynamics of the auxiliary variable $z_o(kT_s)$ can be derived by substituting Eq. (4) into Eq. (15) as follows:

$$z_o((k+1)T_s) = (1-\mathbf{L_o}^T\mathbf{D_{Dn}})z_o(kT_s) + \mathbf{L_o}^T(\mathbf{A_{Dn}}+\mathbf{D_{Dn}}\mathbf{L_o}^T-\mathbf{I_2})\mathbf{x}(kT_s) + \\ \mathbf{L_o}^T\mathbf{B_{Dn}}u(kT_s) + \Delta\tau_{dn}(kT_s) \quad (16)$$

Let us synthesise an observer for the auxiliary variable $z_o(kT_s)$ by substituting the general approximate model of the nominal disturbance variable.

$$\hat{z}_o((k+1)T_s) = (1-\mathbf{L_o}^T\mathbf{D_{Dn}})\hat{z}_o(kT_s) + \mathbf{L_o}^T(\mathbf{A_{Dn}}+\mathbf{D_{Dn}}\mathbf{L_o}^T-\mathbf{I_2})\mathbf{x}(kT_s) + \\ \mathbf{L_o}^T\mathbf{B_{Dn}}u(kT_s) + \hat{\Delta}\tau_{dn}(kT_s) \quad (17)$$

where $\hat{z}_o(kT_s)$ represents the estimated $z_o(kT_s)$ at $kT_s$ seconds and $\hat{\Delta}\tau_{dn}(kT_s)$ is the approximate model for the variation of the nominal disturbance variable between $kT_s$ and $(k+1)T_s$ seconds. From Eq. (14), the first and second order approximations of the variation of the nominal disturbance variable are derived as follows:

$$\hat{\Delta}\tau_{dn}(kT_s) = \hat{\tau}_{dnp}(kT_s) - \hat{\tau}_{dnp}((k-1)T_s) \quad (18)$$

$$\hat{\Delta}\tau_{dn}(kT_s) = \frac{3}{2}\hat{\tau}_{dnp}(kT_s) - 2\hat{\tau}_{dnp}((k-1)T_s) + \frac{3}{2}\hat{\tau}_{dnp}((k-2)T_s) \quad (19)$$

Let us subtract Eq. (17) from Eq. (16) to obtain the error dynamics of the HPDOb.

$$e_o((k+1)T_s) = (1 - \mathbf{L_o}^T \mathbf{D_{Dn}})e_o(kT_s) + \Delta\tau_{dn}(kT_s) - \hat{\Delta}\tau_{dn}(kT_s) \quad (20)$$

where $e_o(kT_s) = z_o(kT_s) - \hat{z}_o(kT_s)$ is the estimation error of the auxiliary variable $z_o(kT_s)$ at $kT_s$ seconds.

Equation (20) shows that the observer gain $\mathbf{L_o}$ should satisfy $|1 - \mathbf{L_o}^T \mathbf{D_{Dn}}| < 1$ to achieve stability in the HPDOb. The estimation error converges to zero asymptotically when the estimation of the nominal disturbance variation is accurately identified, i.e., $\Delta\tau_{dn}(kT_s) = \hat{\Delta}\tau_{dn}(kT_s)$. Nevertheless, this is an impractical requirement in many robust motion control applications. When the estimation of the nominal disturbance variation is bounded, the HPDOb is stable in the sense that the estimation error is uniformly ultimate bounded. The upper bound of disturbance estimation error is directly related to the model of the nominal disturbance variable used in the HPDOb synthesis. The reader is referred to [38] for a more detailed stability analysis for DOb-based digital robust motion control systems.

## IV. SIMULATIONS

To validate the proposed HPDOb, this section presents simulation results. The DOb-based digital robust position

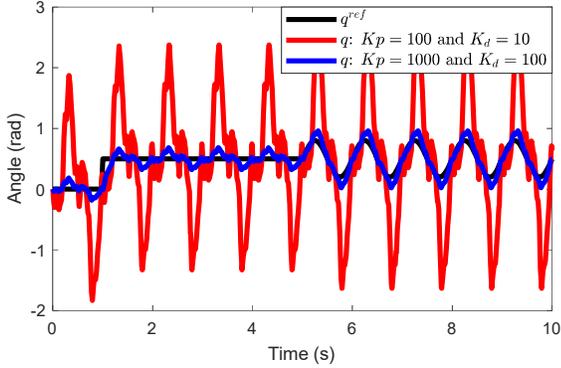

a) Regulation and trajectory tracking control.

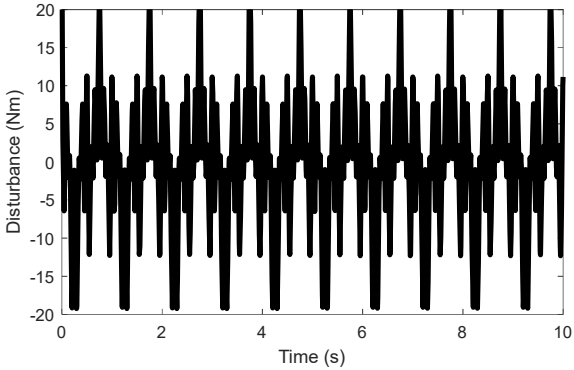

b) External disturbances exerting on the servo system.

Fig. 1: Conventional PD-based position control.

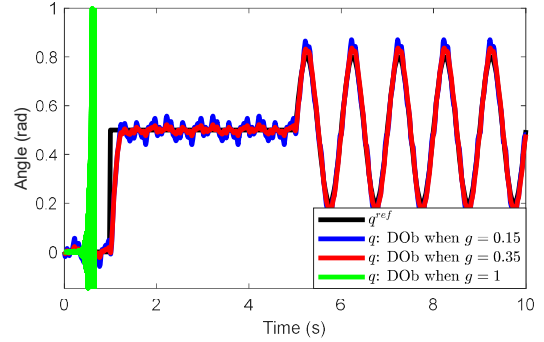

a) Regulation and trajectory tracking control.

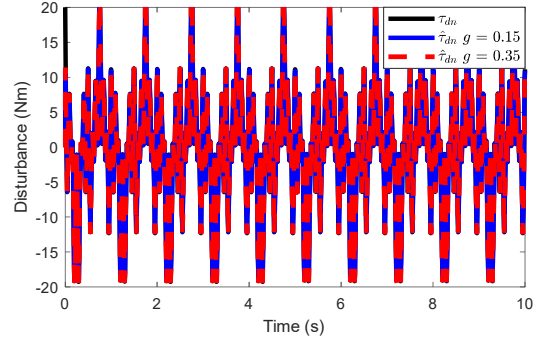

b) Nominal disturbance variable and its estimation.

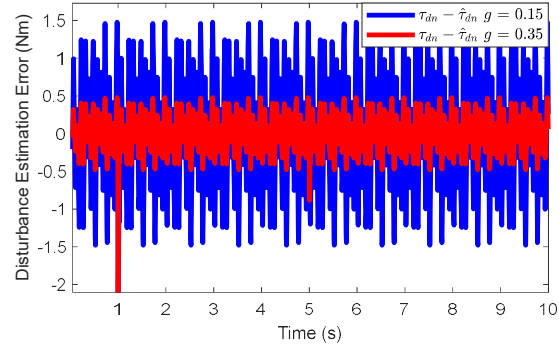

c) Disturbance estimation error.

Fig. 2: Robust position control using the conventional DOb and a PD controller, with the control gains of $K_p = 100$ and $K_d = 10$.

controllers are implemented using the following simulation parameters in MATLAB/SIMULINK: $J_m = 0.125$ kgm², $b_m = 0.045$ Nms/rad, $T_s = 0.1$ms, $K_p = 100$ and $K_d = 10$.

The conventional and proposed high-performance DObs are synthesised using the following observer gain vector.

$$\mathbf{L_*} = \frac{g_*}{|\mathbf{D_{Dn}}|_1}[1 \ 1]^T \quad (21)$$

where $g_*$ is a free control parameter in which $*$ is null, $p$ and $o$ for the conventional, predictor and high-performance observer synthesis, respectively; and $|\mathbf{D_{Dn}}|_1$ is the L1-norm of $\mathbf{D_{Dn}}$.

Let us start simulations with a Proportional and Derivative (PD) based motion control system. Figure 1a illustrates the regulation and trajectory tracking control performance of the PD controller, while the external nonlinear disturbances exerted on the motion controller is illustrated in Fig. 1b. As shown in this figure, the conventional PD controller can provide limited performance when the motion control system is disturbed by external disturbances. Although the performance can be simply improved by increasing the proportional and derivative control gains for 10 times, the trajectory deviation remains relatively large as illustrated in Fig. 1a.

Let us now consider the conventional DOb-based robust motion controller. To improve robustness, the estimated disturbances are fed back using the conventional DOb in the inner-loop, while the position control performance is adjusted using a PD controller with the proportional and derivative gains of $K_p = 100$ and $K_d = 10$ in the outer-loop. Figure 2 illustrates the robust position control simulations when different observer gains are used in the conventional DOb synthesis. Figure 2a shows that the position control performance improves as the control gain of the DOb is increased from 0.15 to 0.35. This performance improvement is achieved, because the DOb can estimate disturbances more accurately when the observer gain is increased as illustrated in Figs. 2b and 2c. However, as shown in Fig. 2a, the stability of the robust motion controller

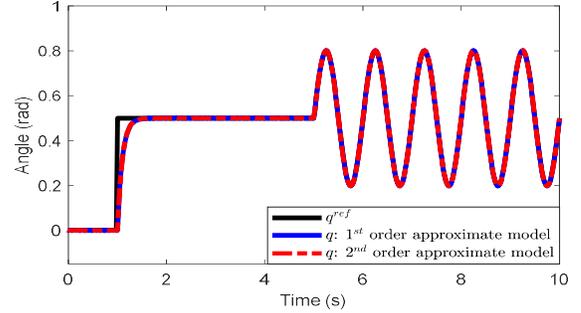

a) Regulation and trajectory tracking control.

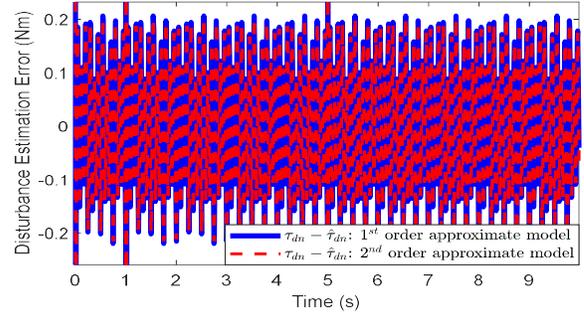

b) Disturbance estimation error.

Fig. 4: Robust position control using the second-order HPDOb with the observer gain of $g_p = g_o = 0.15$ and a PD controller with the control gain of $K_p = 100$ and $Kd = 10$.

deteriorates as the observer gain is increased (see green curve in Fig. 2a). The upper bound on the observer gain can be found using Eq. (9). It is noted that the upper bound on the observer gain changes by the nominal design parameters. The reader is referred to [38, 39] for a detailed explanation of why the stability of the robust motion controller deteriorates as the bandwidth and nominal inertia is increased.

To further improve the robustness of the motion controller, let us employ the proposed HPDOb implemented by using the first order approximate model given in Eq. (18) in the observer synthesis. The position control simulations are illustrated in Fig. 3. As shown in Fig. 3a, the proposed HPDOb allows us to conduct more accurate position control applications compared to the conventional DOb-based robust motion controller. This can be achieved, because the HPDOb can estimate the nominal disturbance variable more accurately than the conventional DOb as illustrated in Figs. 3b and 3c. This can notably benefit robust position control applications in practice.

Last, let us implement the robust motion controller using the HPDOb implemented with the second order approximate model given in Eq. (19) in the observer synthesis. The position control experiments are illustrated in Fig. 4. The proposed second order HPDOb outperforms the conventional DOb and the first order HPDOb, so it can similarly boost the robust position control applications in practice. Nevertheless, although the second order HPDOb provides the most accurate disturbance estimation, the disturbance estimation accuracy and robustness against disturbances are very similar when the first and second order approximate models are employed in the HPDOb synthesis. Thus, high-performance robust motion control applications can be implemented using the first order HPDOb.

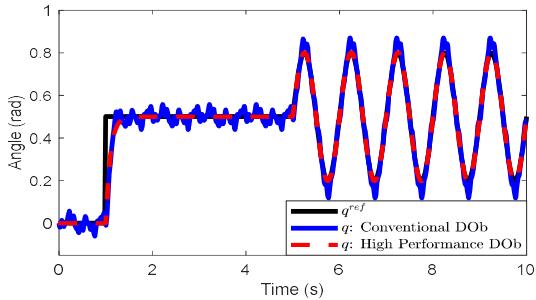

a) Regulation and trajectory tracking control.

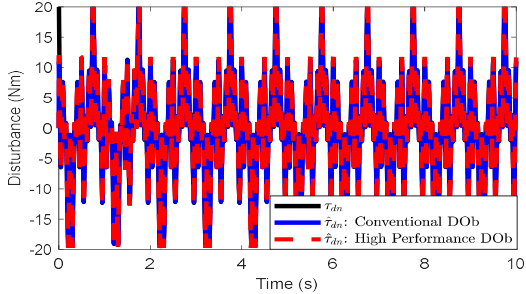

b) Nominal disturbance variable and its estimation.

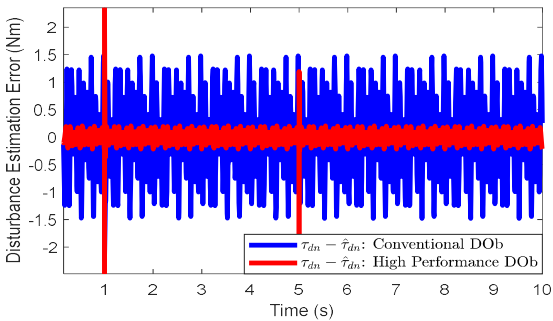

c) Disturbance estimation error.

Fig. 3: Robust position control using the first-order HPDOb with the observer gain of $g_p = g_o = 0.15$ and a PD controller with the control gain of $K_p = 100$ and $Kd = 10$.

## V. Conclusion

This paper has presented a novel high-performance DOb that can estimate disturbances more accurately than the conventional DOb. To clearly represent the dynamic behaviour of the digital robust motion controllers in practice, the conventional and proposed high-performance DObs are analysed and synthesised in the discrete-time domain. While the conventional DOb relies on a zero-order Taylor series expansion of disturbances, the use of higher-order Taylor series expansions in the HPDOb synthesis enables more accurate disturbance estimation. This can significantly improve DOb-based robust motion control applications. For example, disturbances exerting on servo systems can be suppressed more robustly in position control, while interaction forces can be estimated more accurately in sensorless force control. The validity of the proposed HPDOb is verified through simulations of the robust position control problem of a servo system. In future studies, the proposed HPDOb should be verified through experiments and analysis to assess its performance in different DOb-based robust control applications such as precisely compensating time-delay in communication systems.